\documentclass[draftcls]{IEEEtran}
\usepackage[utf8]{inputenc}
\usepackage{mathtools,amssymb}
\usepackage{graphicx}
\usepackage{mathrsfs}
\usepackage[dvipsnames]{xcolor}
\colorlet{bvb}{Red}
\colorlet{wz}{blue}
\colorlet{bvb2}{orange}

\DeclareMathAlphabet\mathbfcal{OMS}{cmsy}{b}{n}

\DeclareMathOperator*{\argmin}{arg\,min}

\DeclareRobustCommand*{\IEEEauthorrefmark}[1]{%
  \raisebox{0pt}[0pt][0pt]{\textsuperscript{\footnotesize #1}}%
}
\begin{document}

\title{Machine Learning for CSI Recreation Based on Prior Knowledge}

\author{
\IEEEauthorblockN{\textit{Brenda Vilas Boas\IEEEauthorrefmark{1}$^,$\IEEEauthorrefmark{2}, Wolfgang Zirwas\IEEEauthorrefmark{1}}, \textit{Martin Haardt\IEEEauthorrefmark{2}}}

\IEEEauthorblockA{\IEEEauthorrefmark{1}Nokia, Germany \\\
\IEEEauthorrefmark{2}Ilmenau University of Technology, Germany \\\
{brenda.vilas\_boas@nokia.com, wolfgang.zirwas@nokia-bell-labs.com, martin.haardt@tu-ilmenau.de} }
% \and 
% \IEEEauthorblockN{\textit{Martin Haardt\IEEEauthorrefmark{2}}}
%\IEEEauthorblockA{\IEEEauthorrefmark{2}Ilmenau University of Technology, Germany \\\ martin.haardt@tu-ilmenau.de }}
}
\maketitle

\begin{abstract}
    Knowledge of channel state information (CSI) is fundamental to many functionalities within the mobile wireless communications systems. With the advance of 
    machine learning (ML) and digital maps, i.e., digital twins, we have a big opportunity to learn the propagation environment and design novel methods to derive and report CSI. 
    In this work, we propose to combine   
    untrained neural networks (UNNs) and conditional generative adversarial networks (cGANs) for MIMO channel recreation based on prior knowledge. 
    The UNNs learn the prior-CSI for some locations which are used to build the input to a cGAN. Based on the prior-CSIs, their locations and the location of the desired channel, the cGAN is trained to output the channel expected at the desired location. This combined approach can be used for low overhead CSI reporting as, after training, we only need to report the desired location. 
    Our results show that our method is successful in modelling the wireless channel and robust to location quantization errors in line of sight conditions. 
    
\end{abstract}


\section{Introduction}

Machine learning (ML) for physical layer applications 
%applications to the physical layer 
are gaining momentum in 
standardization bodies, such as 3GPP and O-RAN
%discussions
~\cite{213gppRel18}. 
Combining ML capabilities with virtual representations of the real world, i.e., a digital 
twin environment%~\cite{19NokiaDigital}
, enables a variety 
of possibilities for wireless network
planning, deployment and management. 
In order to leverage the potential of a digital twin of the environment, 
full knowledge of the channel state information (CSI) is desired such that most of the real propagation effects 
can be represented. 
Here, we propose to combine two ML methods  
%propose to use untrained neural networks (UNNs) 
for channel recreation/estimation which minimize the overall complexity of the neural networks (NNs), reduce training time, and enable low CSI reporting overhead. 
In contrast to state of the art, our solution does not rely on multi-modal data, such as lidar~\cite{19DiasLidar} or environment images~\cite{21RatnanFadeNet}, which allow us to reduce the complexity of our NN architectures.   

% Application: Considering a next generation wireless network in which network planning and deployment is aided by a digital twin. There are a few known CSIs (e.g.: by Untrained Neural Networks (UNNs)) and we aim to estimate/predict the channels of neighboring UEs. We know the location of the prior-CSI and the desired-CSI, our ML leverages those information to estimate the channel at the desired location. Therefore, we say our ML has interpolation-like capabilities.     

% Our contributions: 
% \begin{enumerate}
%     \item cGAN for estimating the desired channel based on prior CSI knowledge and their locations - interpolation. 
%     \item Performance gains for different 'coverage' areas. 
%     \item Present network structure and simulations at IlmProp.
       
% \end{enumerate}

Untrained neural networks (UNNs) were first proposed in~\cite{18LempitskyDIP} to solve inverse problems, such as denoising. The term `untrained' refers to the method characteristic of avoiding a huge data collection phase as the updates of the gradient descent is for
%take into account 
a single image measurement.  
%The work in~\cite{18LempitskyDIP} has proposed to use neural network architectures based on convolution operations to fit a single image measurement, instead of a data collection. This approach is refereed as untrained neural networks (UNNs) 
%were first proposed in~\cite{18LempitskyDIP} where the term untrained relates to the fact that a huge data collection for training is not needed. Instead, the neural network can be fitted to a single data sample. Therefore, 
%as the gradient descent updates move from the training phase to the inference phase. The solution of inverse problems, such as denoising, is doable because the structure of deep convolutional networks act as a prior to an image-like signal~\cite{18LempitskyDIP}. 
The \textit{deep decoder} architecture as  proposed in~\cite{18HeckelDeep} simplifies the structure of a UNN, making it underparameterized.
%\textit{deep decoder} architecture for UNNs. Here, we use an architecture based on the deep decoder structure for the channel estimation task of the first ML instance. 
For wireless communications, this means we can fit a UNN to directly estimate the wireless channel based on a small noisy measurement campaign, i.e., a few time snapshots, without the need of noiseless labels.
The work in ~\cite{20BaleviUNN} has proposed the use of UNNs for MIMO channel estimation under pilot contamination. 
Despite the limitation to statistical channel models, UNNs could reduce the noise level of the measured signal.

The simplicity of UNNs comes at the cost of lack of generalization. 
Since there is no dataset collection for weights update, iterating the gradient descent is always needed when a new set of channel measurements is acquired. 
%Different from UNNs, generative adversarial networks (GANs) need a big data collection and training phases which lead the generator neural network to mimic the data distribution~\cite{14GoodfellowGAN}, generalization capability. 
%For wireless mobile radio systems, GANs are often concerned with physical layer issues like channel modeling  
%and data augmentation~\cite{19YangGANmodeling,19OsheaVenc}. 
%Recently,~\cite{20BaleviHigh} has proposed to find low dimension channel representations based on GANs which were estimating the wireless channel. 
%Conditional GANs (cGANs) provide some prior-knowledge to their generator and discriminator NNs which should ease the mapping task~\cite{14MirzaConditional}. 
%A cGAN is used in~\cite{18LiGANcov} to estimate the millimeter wave (mm-Wave) \textit{virtual} covariance channel matrix based on prior knowledge of a training sequence.       
%A cGAN and a variational autoencoder (VAE) GAN are used in~\cite{18YeChannelE2E,19SmithCommDensity}, but in a context of end-to-end learning where the final objective is to predict the transmitted symbols, not the wireless channel. 
In our recent work~\cite{21BoasTwo}, we have proposed to used cGAN for channel estimation in MIMO arrays with mixed radio frequency chains where part of the array had antenna elements turned-off. Our results demonstrated the good generalization capability of cGANs.    

Motivated by the generalization capabilities of cGAN and the underparameterization of UNNs, we propose to combine them for MIMO channel estimation/prediction within a propagation area.
%Here, we aim to combine the underparameterization of UNNs with the generalization of cGANs for MIMO channel estimation. 
Hence, the UNNs are used to generate prior CSI for a set of locations. 
Then, the cGAN uses the prior-CSIs together with their locations to recreate the CSI in a desired location.     
After deriving the weights for all the ML models, only the target location needs to be reported. 
Therefore, our solution enables low CSI reporting overhead between user equipment (UE) and base station (BS).  
Moreover, our approach can be used to 
add to the digital twin the
%populate digital twin environments with 
small scaling fading characteristics of the wireless channels.

In this paper, 
Section~\ref{sec:scenario} presents our geometrical propagation environment and the channels considered, Section~\ref{sec:method} introduces our proposed method, Section~\ref{sec:UNNs} presents details about our UNN for prior knowledge CSI estimation, Section~\ref{sec:cGAN} shows the processing performed at the cGAN for CSI recreation using location and prior-CSI, 
Section~\ref{sec:results} presents our results, and Section~\ref{sec:conclusion} concludes our paper.

Regarding the notation, $a$, $\mathbf{a}$, $\mathbf{A}$ and $\mathbfcal{A}$ represents, respectively, scalars, column vectors, matrices and $D$-dimensional tensors. The superscript $^T$, denotes transposition. 
For a tensor $\mathbfcal{A} \in \mathbb{C}^{M_1 \times M_2 \times \dots M_D}$, $M_d$ refers to the tensor dimension on the $d^\mathrm{th}$ mode.
A $d$-mode unfolding of a tensor is written as $[\mathbfcal{A}]_{(d)} \in \mathbb{C}^{M_d \times M_{d+1} \dots M_D M_1 \dots M_{d-1}}$ where all $d$-mode vectors are aligned as columns of a matrix. The $d$-mode vectors of $\mathbfcal{A}$ are obtained by 
varying the $d^\mathrm{th}$ index from $1$ to $M_d$ and keeping all other indices fixed.
Moreover, $\mathbfcal{A} \times_d \mathbf{U}$ is the $d$-mode product between a $D$-way tensor $\mathbfcal{A} \in \mathbb{C}^{M_1 \times M_{2} \dots \times M_D}$ %of size $M_d$ along the mode $d=1,2,\dots, M_D$, 
and a matrix $\mathbf{U} \in \mathbb{C}^{J \times M_d}$. The $d$-mode product is computed by multiplying $\mathbf{U}$ with
all $d$-mode vectors of
$\mathbfcal{A}$.
In addition, $\mathbfcal{A} \sqcup_d \mathbfcal{B}$ denotes the concatenation of $\mathbfcal{A}$ and $\mathbfcal{B}$ among the $d^\mathrm{th}$ mode. The concatenation $\sqcup_d$ operation also applies to matrices.

\section{Propagation Environment}
\label{sec:scenario}

\begin{figure}[!tb]
    \centering
    \includegraphics[width=0.7\columnwidth]{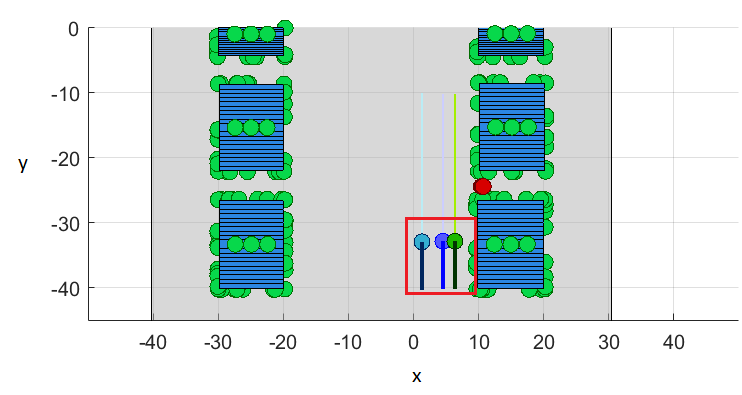}
    \caption{Propagation scenario simulated in IlmProp, a geometric based channel simulator developed at Ilmenau University of Technology. The BS is fixed and represented by a red circle. The displayed UEs and trajectories are used to compute the prior-CSIs. The red square represents the study area ($100~\mathrm{m}^2$), to which we aim to reconstruct the CSIs.}
    \label{fig:IlmProp}
\end{figure}

In this work, we consider an urban environment with a fixed base station (BS) equipped with an uniform rectangular array (URA) containing $N_\mathrm{ant}$ antenna elements, moving user equipment (UEs) with single antennas, operating with $N_\mathrm{sub}$ OFDM subcarriers, and collecting $N_\mathrm{sp}$ time snapshots. 
This scenario was modeled with IlmProp, a geometry based channel simulator developed at Ilmenau University of Technology~\cite{05GaldoGeometry}.  
Figure~\ref{fig:IlmProp} presents the urban environment with the BS represented by a red circle, buildings in blue squares, scatters in green circles, and three UEs moving in a linear trajectory towards the BS. 

% For wireless channel recreation, we define two types of channel state information (CSI): the prior-CSI $\mathbfcal{H}_p \in \mathbb{C}^{N_\mathrm{ant} \times N_\mathrm{sub} \times N_\mathrm{sp}}$ 
% and the recreated-CSI $\mathbfcal{H}_r \in \mathbb{C}^{N_\mathrm{UE} \times N_\mathrm{ant} \times N_\mathrm{sub}}$ where $N_\mathrm{UE}$ is the number of UEs considered. The prior-CSI is a wireless channel that we know a priory for some locations. The reconstructed-CSI is the wireless channel we aim to estimate based on the knowledge of its location and the selected prior-CSIs. The three UEs shown in Figure~\ref{fig:IlmProp} are used to generate the prior-CSIs, whereas the red square indicates the study area where we collect channels for CSI reconstruction. 

From IlmProp, we collect the channels used as ground truth values
%The channels generated at IlmProp 
$\mathbfcal{H}_\mathrm{sim}^C \in \mathbb{C}^{N_\mathrm{sp} \times N_\mathrm{sub} \times N_\mathrm{ant}}$ 
%used as ground truth values 
and their relative locations $\mathbf{\Gamma} = \{ \mathbf{x}, \mathbf{y}, \mathbf{z} \} \in \mathbb{R}^{N_\mathrm{sp} \times 3}$ to the BS.  
The three UEs shown in Figure~\ref{fig:IlmProp} are used to derive the noisy channel measurements $\mathbfcal{H}_\mathrm{mes}^C \in \mathbb{C}^{N_\mathrm{sp} \times N_\mathrm{sub} \times N_\mathrm{ant}}$ as %prior-CSIs, 
\begin{equation}
    \mathbfcal{H}_\mathrm{mes}^C = \mathbfcal{H}_\mathrm{sim}^C + \mathbfcal{N}, 
    \label{eq:mes}
\end{equation}
where $\mathbfcal{N} \in \mathbb{C}^{N_\mathrm{sp} \times N_\mathrm{sub} \times N_\mathrm{ant}}$ is a zero mean circularly symmetric complex Gaussian noise process.
The $\mathbfcal{H}_\mathrm{mes}^C$ are further used to estimate the prior-CSIs.    
Moreover, the red square indicates the study area where the $\mathbfcal{H}_\mathrm{sim}^C$ are collected for CSI recreation.

%Both $\mathbfcal{H}_p \in \mathbb{C}^{N_\mathrm{ant} \times N_\mathrm{sub} \times N_\mathrm{sp}}$ and $\mathbfcal{H}_r \in \mathbb{C}^{N_\mathrm{ant} \times N_\mathrm{sub} \times N_\mathrm{sp}}$. are ground truth values generated from IlmProp. In real life, those channels could be acquired by channel measurements campaigns or drive tests. 

\setlength{\textfloatsep}{1\baselineskip plus 0.2\baselineskip minus 0.2\baselineskip}

\section{CSI recreation with prior knowledge}
\label{sec:method}

\begin{figure}
    \centering
    \includegraphics[width=0.85\columnwidth]{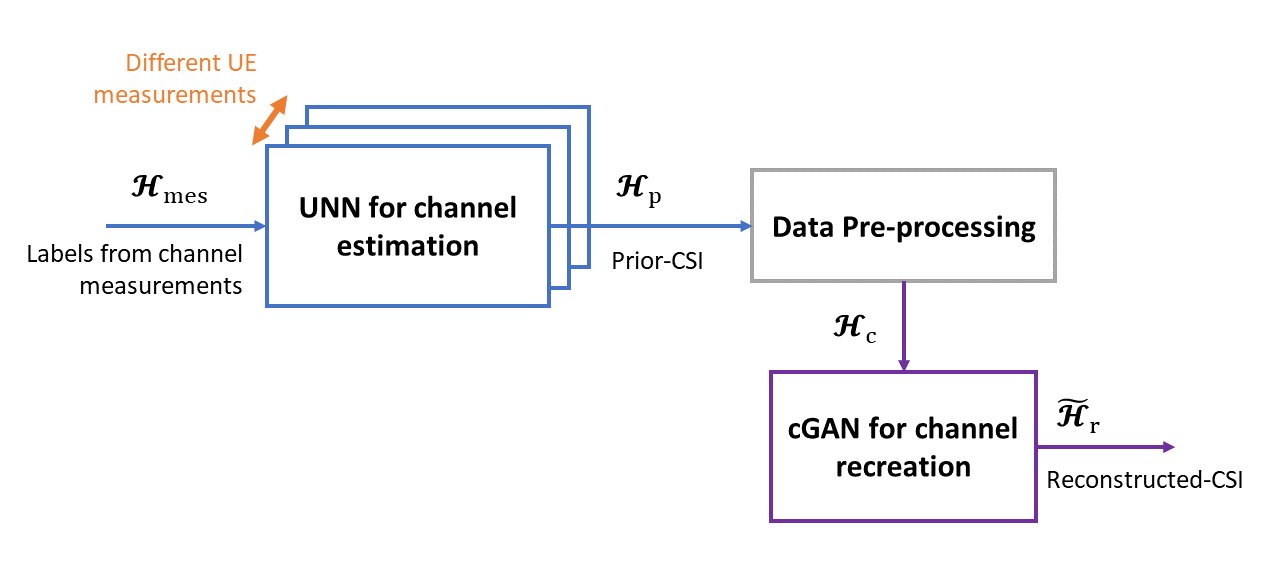}
    \caption{Schematic of our proposed ML solution for CSI recreation combining channel measurements and simulations. % to populate a digital twin environment that is further leveraged for wireless system planing and deployment. 
    In blue, we present the first ML part where the UNNs are iterated to perform channel estimation. Each UE measurement campaign derive one UNN that represents $N_\mathrm{sp}$ locations. In purple, we show the second ML part where a cGAN is trained with knowledge of the prior-CSIs, the simulated desired channels and their locations. The cGAN output is the recreated desired channel.}
    \label{fig:system}
\end{figure}

Here, we propose a ML framework to recreate CSI in a desired location based on prior knowledge of CSI at neighboring UEs.  
Figure~\ref{fig:system} summarizes our proposed method for CSI recreation which we divide in two parts, the first in blue and the second in purple. 
The first ML instance aims to find the prior-CSI $\mathbfcal{H}_\mathrm{p} \in \mathbb{R}^{N_\mathrm{sp} \times N_\mathrm{sub} \times 2N_\mathrm{ant}}$ 
based on the measured channels $\mathbfcal{H}_\mathrm{mes}^C$. We employ a %untrained neural network (UNN) 
UNN for this purpose where each UNN estimates 
$\mathbfcal{H}_\mathrm{p}$ the channel of a single UE over multiple time snapshots. 
%Each time snapshot $N_\mathrm{sp}$ measurement is referred as an instantaneous measurement 
%for an $i^\mathrm{th}$ UE at position $\mathrm{\textbf{LOC}}_{p,i} \in \mathbb{R}^{1 \times 3}$ where $i=[1, 2, \ldots, N_\mathrm{sp}]$. 
%If UNNs are used to estimate the prior-CSI of multiple $N_\mathrm{UE}$ users, then there are prior-CSIs available in $N_\mathrm{UE}N_\mathrm{sp}$ positions. %Therefore, more prior-CSIs increase the 
Even though UNNs are low complex structures, deriving one UNN model for each possible location in a propagation environment is unfeasible. 
Therefore, we propose to use a second ML instance based on a cGAN due to its generalization capabilities.   
The second ML instance is trained to compute the recreated-CSI $\mathbfcal{H}_\mathrm{r} \in \mathbb{R}^{(S+1)  \times N_\mathrm{sub} \times (2N_\mathrm{ant}+1)}$ in the targeted location 
$\mathbf{\Gamma}_\mathrm{r} \in \mathbb{R}^{1 \times 3}$
%$\mathrm{\textbf{LOC}}_\mathrm{r} \in \mathbb{R}^{1 \times 3}$
%, where $N_\mathrm{UE}$ is the number of UEs considered at the second ML output, 
based on the knowledge of a sub-set of $S$ selected prior-CSIs and their respective locations
$\mathbf{\Gamma}_\mathrm{p} \in \mathbb{R}^{S \times 3}$.
%$\{\mathbf{H}_{\mathrm{p},1},\ldots, \mathbf{H}_{\mathrm{p},M-1}\} \subset  \mathbfcal{H}_\mathrm{p}^R$, and their respective locations 
%$\{ \mathrm{\textbf{LOC}}_{\mathrm{p},1},\ldots,~\mathrm{\textbf{LOC}}_{\mathrm{p},M-1} \} \subset \mathrm{\textbf{LOC}}_{\mathrm{p}}$. 
%For the task of channel estimation based on prior knowledge,
%we use a conditional generative adversarial network (cGAN). 

Since UNNs do not need `labels' to find their best weights, we can perform a small measurement campaign and use the UNN-estimated channels as conditional input to the cGAN. 
In a `day-zero' operation where not many CSI measurements are available, the cGAN can be trained with target channels derived from simulations 
or a digital twin
and the prior-CSIs from the UNNs are responsible to adjust the model to real propagation conditions. 
In the long run, we could update the cGAN model based on collected real world measurements. 
In this scenario, the availability of priors at the conditional input reduce the complexity of the NN structure and its training time.   
In the following sections, we explain in details how each part of the algorithm is trained.
%where $i=[1, 2, \ldots, N_\mathrm{sp}]$ is the set in which the prior-CSIs are selected, and the targeted location $\mathrm{LOC}_\mathrm{r}$ where we aim to recreate $\mathbfcal{H}_\mathrm{r}$. 
%For the task of channel estimation performed by the first ML instance, we have selected a untrained neural network (UNN). While for channel recreation based on prior knowledge, we have selected a conditional generative adversarial network (cGAN). In the following sections we explain in details how each part of the algorithm operates.  

Due to the low computational complexity of UNNs, the UE can derive the UNN weights and send it to the BS. 
Then, the BS is able to reconstruct the prior-CSIs and can train a cGAN to recreate the CSI at a desired location, which is different from the prior-CSI locations. 
After deriving the cGAN optimal weights, the BS can send the cGAN model to the UE for the purpose of reliability. Hence, the UE is able to identify when the BS will fail on its CSI recreation and may trigger a correction procedure. 

\section{Estimation of prior-CSI with UNNs}
\label{sec:UNNs}

The underparametrization of deep decoder~\cite{18HeckelDeep} and 
its capability to optimize noisy measurements
%their optimization directly over noisy measurements 
has motivated us to employ UNNs as our channel estimator for prior-CSIs. 
Since we do not need the true-labels, the channel measurements can be acquired during measurement campaigns which will then represent the real world propagation environment for specific locations. 
%Then, representing the real-world propagation environment.   
The following subsections present the data pre-processing, the UNN architecture and how the gradient descent is used to update the UNN weights.           

\begin{figure}[!tb]
    \centering
    \includegraphics[width=0.75\columnwidth]{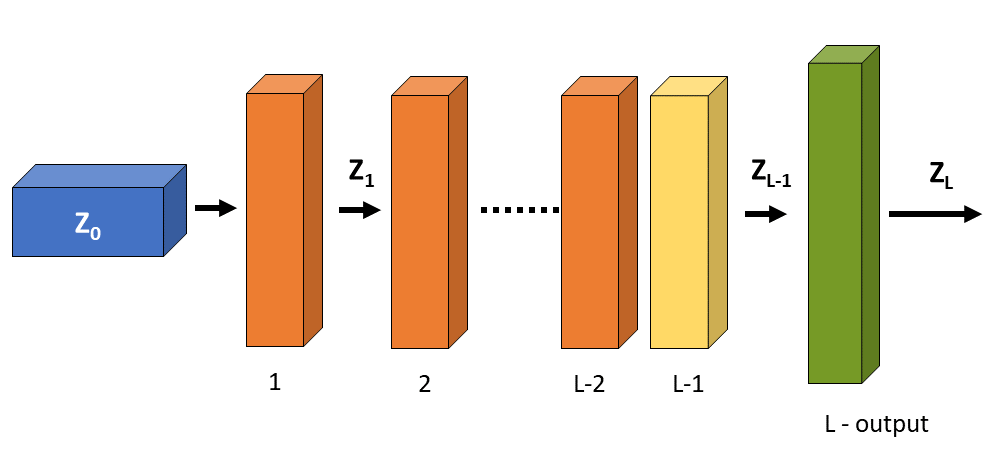}
    \caption{General layer structure of a UNN $P$ used to estimate the prior-CSIs $\mathbfcal{H}_\mathrm{p} = P(\mathbfcal{K}^*, \mathbfcal{Z}_0)$. There are $L$ layers, $L-2$ inner layers in orange, one pre-output layer in yellow, and one output layer in olive. In blue, we represent $\mathbfcal{Z}_0$ the random input tensor.}
    \label{fig:unn}
\end{figure}

\subsection{Data pre-processing for UNN}

The input signal to an UNN is a random noise seed $\mathbfcal{Z}_0 \in \mathbb{R}^{k_1 \times b \times c}$, where $b = N_\mathrm{sp}/2^{L-2}$, $c = N_\mathrm{sub}/2^{L-2}$,~$k_1$ is the number of filters in the first layer, and $L$ is the number of layers. The input tensor %random seed
$\mathbfcal{Z}_0$ is drawn from a uniform distribution $U(-a,+a)$ defined on the interval $[-a, +a]$ and kept fixed during the iterations to update the gradient descent.   
The measured channel $\mathbfcal{H}_\mathrm{mes}^C$ from Equation~\ref{eq:mes} is preprocessed as 
\begin{itemize}
    \item Each time snapshot within $\mathbfcal{H}_\mathrm{mes}^C$ is normalized by its Frobenius norm, and then multiplied by a scaling factor to ease convergence.
    \item $\mathbfcal{H}_\mathrm{mes}^C \in \mathbb{C}^{N_\mathrm{sp} \times N_\mathrm{sub} \times N_\mathrm{ant}}$ is rearranged by concatenating $\mathfrak{Re} \{ \mathbfcal{H}_\mathrm{mes}^C \}$ and $\mathfrak{Im} \{ \mathbfcal{H}_\mathrm{mes}^C \}$ in the dimension corresponding to the antenna elements. 
\end{itemize}
After those operations, $\mathbfcal{H}_\mathrm{mes} \in \mathbb{R}^{N_\mathrm{sp} \times N_\mathrm{sub} \times 2N_\mathrm{ant}}$ is directly used to compute the cost function. 

\subsection{UNN architecture}

A UNN is a composition of $L$ layers where there are
%According to the deep decoder architecture, the architecture has three layers types which we refer as: 
$(L-2)$ inner layers, one pre-output layer $(L-1)$ and one output layer $(L)$, according to the \textit{deep decoder} architecture~\cite{18HeckelDeep}.
Figure~\ref{fig:unn} shows a generic organization of those layers, the random noise seed $\mathbfcal{Z}_0$ in blue, the inner layers in orange, the pre-output layer in yellow, and the output layer in olive. 
All the layer types contain convolutional filters $\mathbfcal{W}_l \in \mathbb{R}^{1 \times 1 \times k_{l-1} \times k_{l}}$ where $l = \{1, 2, \ldots L \}$, $k_{l-1}$ and $k_{l}$ are hyper-parameters which define the number of filters on the respective
$(l-1)^\mathrm{th}$ and $l^\mathrm{th}$ layers. However, the types of layers differ
with respect to the upsampling computation and the operation of batch normalization ($\mathrm{BatchNorm}$) operation~\cite{15IoffeBatch}. 
%All the layer types contain convolutional filters $\mathbfcal{W}_l$ with $l = \{1, 2, \ldots L \}$, but they differ with respect to the upsampling computation and the batch normalization ($\mathrm{BatchNorm}$) operation~\cite{15IoffeBatch}.
The inner layers contain linear and non-linear operations.
First, there is a convolutional filter $\mathbfcal{W}_l$ which weights are updated by the gradient descent. 
%$\in \mathbb{R}^{1 \times 1 \times k_l \times k_{l+1}}$ where $k_l$ and $k_{l+1}$ are the number of filters in the $l^\mathrm{th}$ and the $(l+1)^\mathrm{th}$ layer, respectively. 
Second, there is a fixed bilinear upsampling operation, where $\mathbf{A}_l \in \mathbb{R}^{{2^l}b \times 2^{l-1}b}$ and  $\mathbf{C}_l \in \mathbb{R}^{2^{l}c \times 2^{l-1}c}$ are the
%one dimensional 
linear upsampling matrices in the subcarrier and time snapshots dimensions, respectively. 
Third, the rectifier linear unit (ReLu) activation function is applied, and a batch normalization is 
computed with trainable parameters $\mathbf{R}_l = [ \boldsymbol{\gamma}_l, \boldsymbol{\beta}_l ] \in \mathbb{R}^{k_l \times 2}$, 
where $\boldsymbol{\gamma}_l \in \mathbb{R}^{k_l \times 1}$
and $\boldsymbol{\beta}_l \in \mathbb{R}^{k_l \times 1}$ 
are the mean and variance correction factors, respectively, for the coefficients in the $l^\mathrm{th}$ layer. 
For example, the output of the first inner layer $\mathbfcal{Z}_1 $ can be written as
\begin{equation}
    \mathbfcal{Z}_1 = \mathrm{BatchNorm} (\mathrm{ReLu}(\mathbfcal{Z}_0 \times_3 [\mathbfcal{W}_1]_{(4)} \times_1 \mathbf{A}_1 \times_2 \mathbf{C}_1^T)),
\end{equation}
where %we further simplify for $N_\mathrm{sp}=N_\mathrm{sub}$ which lead to $\mathbf{C}_l=\mathbf{A}_l^T$, and
$[\mathbfcal{W}_1]_{(4)}$ is the $4$-mode unfolding of the convolutional filters operating at the antenna elements dimension.  
The pre-output layer differs from the inner layers because it does not apply upsampling. Hence, it can be written as
\begin{equation}
    \mathbfcal{Z}_{L-1} = \mathrm{BatchNorm} (\mathrm{ReLu}(\mathbfcal{Z}_{L-2} \times_3 [\mathbfcal{W}_{L-1}]_{(4)})).
\end{equation}
Next, the output layer is used to adjust the number of filters of the pre-output layer to the size expected at the output $k_L = 2N_\mathrm{ant}$ as
\begin{equation}
    \mathbfcal{Z}_{L} = \mathrm{TanH}(\mathbfcal{Z}_{L-1} \times_3 [\mathbfcal{W}_{L}]_{(4)}),
\end{equation}
where $\mathbfcal{W}_{L} \in \mathbb{R}^{1 \times 1 \times k_{l-1} \times 2N_\mathrm{ant}}$, and $\mathrm{TanH}$ is the hyperbolic tangent activation function.
Since the upsamplig operations are pre-defined, the trainable parameters relates to the convolutional filters
$\mathbfcal{W}_{l}$ and the regularization parameters $\mathbf{R}_l$ of the batch normalization operation.
Therefore, $\mathbfcal{K}_l = \{\mathbfcal{W}_{l}, \mathbf{R}_l\}$ is the set of trainable parameters of the $l^\mathrm{th}$ layer, 
and $\mathbfcal{K}$ refers to all trainable parameters of the $L$ layers.

\subsection{Updating the weights of an UNN}

Here, we refer to the UNN as a model $P:~\mathbb{R}^N~ \rightarrow~\mathbb{R}^{N_\mathrm{sub}N_\mathrm{sp}2N_\mathrm{ant}}$ where
$N<N_\mathrm{sub}N_\mathrm{sp}2N_\mathrm{ant}$ is the total number of parameters. The UNN $P$ performs the mapping operation $\mathbfcal{Z}_L  = P(\mathbfcal{K},\mathbfcal{Z}_0)$, where $\mathbfcal{Z}_0$ is the random noise seed, and $\mathbfcal{K}$ is the tensor of weights that represents all the UNN trainable parameters. % $L$ convolutional filters and $(L-1)$ batch normalization parameters.

The cost function %to compute the gradient descent and update $\mathbfcal{K}$ 
is the mean square error (MSE), calculated as
\begin{equation}
    \mathcal{L}(\mathbfcal{K}) = \mathbb{E} \{\parallel P(\mathbfcal{K}, \mathbfcal{Z}_0) - \mathbfcal{H}_\mathrm{mes} \parallel^2_F \}.
\end{equation}
%\begin{equation}
 %   \mathcal{L}(\mathbfcal{K}) = \parallel P(\mathbfcal{K}, \mathbfcal{Z}_0) - \mathbfcal{H}_\mathrm{mes} \parallel^2_2.
%\end{equation}
The gradient descent is updated as in supervised learning, performing $I$ gradient iterations until the optimum parameters are found, such that
\begin{equation}
    \mathbfcal{K}^* = \underset{\mathbfcal{K}}{\argmin}~ \mathcal{L}(\mathbfcal{K}), ~ \mathrm{and} ~
    \mathbfcal{H}_\mathrm{p} = P(\mathbfcal{K}^*, \mathbfcal{Z}_0)
\end{equation}
is the channel estimation of the prior-CSI. 
From the loss function, we observe that the prior-CSI $\mathbfcal{H}_\mathrm{p}$ derived by the UNN $P$ is specific to $\mathbfcal{H}_\mathrm{mes}$. Hence, the model $P$ does not directly generalize for other channels, it is specific to the $\mathbfcal{H}_\mathrm{mes}$ considered during gradient updates.     

\section{CSI-recreation with cGAN}
\label{sec:cGAN}

\begin{figure}[!bt]
\centering
\includegraphics[width=0.85\columnwidth]{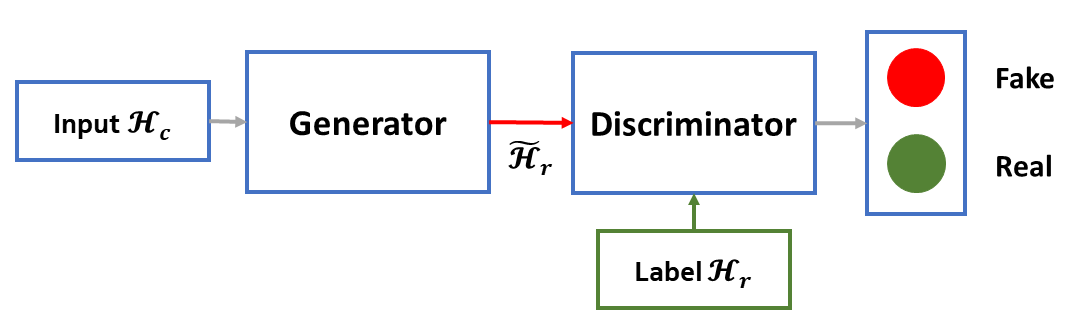}
\caption{Conditional GAN, two NNs play a minmax game where the generator tries to fool the discriminator. %Both NNs have knowledge of the prior information $\mathbfcal{H}_\mathrm{c}$.
The discriminator should classify  $\mathbfcal{\tilde{H}}_\mathrm{r}$ %For the discriminator, $[\mathbf{\tilde{H}}_g,\mathbf{H}_\mathrm{ce}]$ %the output of the generator $\mathbf{\tilde{H}}_g$ concatenated with its correspondent $\mathbf{H}_\mathrm{ce}$ 
%should be classified 
as a fake sample, while $\mathbfcal{H}_\mathrm{r}$ is classified as a real sample. 
The generator fools the discriminator when $\mathbfcal{\tilde{H}}_\mathrm{r}$ is classified as real.} %~\cite{14GoodfellowGAN}.}
\label{fig:cgan}
\end{figure}

In our previous work~\cite{21BoasTwo}, we have applied cGAN for the purpose of channel estimation within mixed-resolution radio-frequency chains. Here, we use the same operational principle of image to image translation~\cite{17IsolaPix}. %to estimate missing data points. 
However, we aim 
to reconstruct CSI based on the knowledge of its location and a set of prior-CSIs. Different from UNNs, cGAN requires data collection and training. Nonetheless, it has great generalization capabilities~\cite{21BoasTwo}.  
%Figure~\ref{fig:cgan} depicts the interplay between the two neural networks (NNs) in the cGAN training phase: the generator and the discriminator. %a cGAN training. 
In the following subsections, we present the dataset preprocessing, our cGAN architecture, and the adversarial training.

\subsection{Dataset preprocesing for cGAN}
\label{sub:datagan}

In this section, we present how we construct the signals to train the cGAN: the conditional input $\mathbfcal{H}_\mathrm{c}$, the label $\mathbfcal{H}_\mathrm{r}$, and the generator output $\tilde{\mathbfcal{H}}_\mathrm{r}$.

The conditional input to our cGAN is derived from the prior-CSIs $\mathbfcal{H}_\mathrm{p}$, their locations
$\mathbf{\Gamma}_\mathrm{p}$, and the target location $\mathbf{\Gamma}_\mathrm{r}$ where we aim to recover the CSI of a certain UE. 
Each $\mathbfcal{H}_\mathrm{p} \in \mathbb{R}^{N_\mathrm{sub} \times N_\mathrm{sp} \times 2N_\mathrm{ant}}$ estimated by a UNN has CSI for $N_\mathrm{sp}$ different locations. 
Therefore, if $N_\mathrm{UE}$ UNNs are used to estimate the prior-CSIs $\mathbfcal{H}_\mathrm{p}^{N_\mathrm{UE}} \in \mathbb{R}^{N_\mathrm{UE}N_\mathrm{sp} \times N_\mathrm{sub} \times 2N_\mathrm{ant}}$, there are  $N_\mathrm{UE}N_\mathrm{sp}$ CSI-location pairs $\{ \mathbf{H}_{\mathrm{p}j}, \mathbf{\Gamma}_{\mathrm{p}j} \}$ 
available, where $\mathbf{H}_{\mathrm{p}j} = \mathbfcal{H}_\mathrm{p}^{N_\mathrm{UE}}(j,:,:)$ %$\in \mathbb{R}^{N_\mathrm{UE}N_\mathrm{sp} \times N_\mathrm{sub} \times 2N_\mathrm{ant}}$ 
and $j=\{1, 2, \dots N_\mathrm{UE} N_\mathrm{sp} \}$.
%is the concatenation of all $N_\mathrm{UE}$ UNNs in the first mode. 
From the available CSI-location pairs, a sub-set of $S$ CSI-location pairs is selected according to their minimum Euclidean distance to $\mathbf{\Gamma}_\mathrm{r}$. 
The $S$ selected prior-CSIs $\mathbfcal{H}_\mathrm{p}^S \in \mathbb{R}^{S \times N_\mathrm{sub} \times 2N_\mathrm{ant}}$ are concatenated in the first dimension and ordered according to the minimum Euclidean distance to the target location $\mathbf{\Gamma}_\mathrm{r}$. 
The target location vector $\mathbf{\Gamma}_\mathrm{r} \in \mathbb{R}^{1 \times 3}$ and the prior location matrix $\mathbf{\Gamma}_\mathrm{p}^S \in \mathbb{R}^{S \times 3}$ are extended by repeating their coordinates until $\mathbf{\Gamma}_\mathrm{p}^S \in \mathbb{R}^{S \times N_\mathrm{sub}}$ and $\mathbf{\Gamma}_\mathrm{r} \in \mathbb{R}^{1 \times N_\mathrm{sub}}$. 
Hence, the complete location matrix is formed as $\mathbf{H}_\mathrm{LOC} = [\mathbf{\Gamma}_\mathrm{r} \sqcup_1 \mathbf{\Gamma}_\mathrm{p}^S] \in \mathbb{R}^{(S+1) \times N_\mathrm{sub}}$. 
Finally, the conditional input to the cGAN is constructed as
\begin{equation}
    \mathbfcal{H}_c = [ \mathbf{H}_N \sqcup_1 \mathbfcal{H}_\mathrm{p}^S \sqcup_3 \mathbf{H}_\mathrm{LOC}] \in \mathbb{R}^{(S+1) \times N_\mathrm{sub} \times (2N_\mathrm{ant}+1)},
\end{equation}
where $\mathbf{H}_N \in \mathbb{R}^{N_\mathrm{sub} \times 2N_\mathrm{ant}}$ is a matrix of random values drawn from a Gaussian distribution. The desired channel $\mathbf{H}_r$ is recreated in the $\mathbf{H}_N$ position at the generator output.  

Ideally, the true recreated CSI $\mathbfcal{H}_\mathrm{r}~\in~ \mathbb{R}^{(S+1) \times N_\mathrm{sub} \times (2N_\mathrm{ant}+1)}$ is found in the output of the cGAN. Hence, each $d$ time snapshot in $\mathbf{H}_\mathrm{sim}(d) = \mathbfcal{H}_\mathrm{sim}^C(d,:,:)$ is used as ground truth value for $\mathbf{H}_r$, the CSI reconstructed at the target location $\mathbf{\Gamma}_\mathrm{r}$.
%$\mathrm{\textbf{LOC}}_\mathrm{r}$.
The pre-processing of the labels for cGAN include
\begin{itemize}
    \item Normalization of $\mathbf{H}_\mathrm{sim}(d) \in \mathbb{C}^{N_\mathrm{sub} \times N_\mathrm{ant}}$ by its Frobenius norm, and multiplication by a scaling factor.
    
    \item $\mathbf{H}_r = [  \mathfrak{Re}\{\mathbf{H}_\mathrm{sim}(d)\} \sqcup_2 \mathfrak{Im}\{\mathbf{H}_\mathrm{sim}(d)\} ] \in \mathbb{R}^{N_\mathrm{sub} \times 2N_\mathrm{ant}}$ is the target real valued CSI at location
    $\mathbf{\Gamma}_\mathrm{r}$.
    %$\mathrm{\textbf{LOC}}_\mathrm{r}$. 
\end{itemize}
Finally, the label is constructed as
\begin{equation}
    \mathbfcal{H}_\mathrm{r} = [ \mathbf{H}_r \sqcup_1 \mathbfcal{H}_\mathrm{p}^S \sqcup_3 \mathbf{H}_\mathrm{LOC}] \in \mathbb{R}^{(S+1) \times N_\mathrm{sub} \times (2N_\mathrm{ant}+1)},
\end{equation}
where the prior-CSIs $\mathbfcal{H}_\mathrm{p}^S$, the location matrix $\mathbf{H}_\mathrm{LOC}$ and the recreated CSI $\mathbf{H}_r$ form the desired output. 
We refer to the generator output as $\tilde{\mathbfcal{H}}_\mathrm{r} = G(\mathbfcal{H}_\mathrm{c}) \in \mathbb{R}^{(S+1) \times N_\mathrm{sub} \times (2N_\mathrm{ant}+1)}$, where $G$ is the generator mapping function that tries to approximate the label $\mathbfcal{H}_\mathrm{r}$.  The discriminator $D$ is a classifier for which the inputs and labels are, respectively, %$[\mathbfcal{H}_\mathrm{c} \sqcup_3 \mathbfcal{H}_\mathrm{r}] \rightarrow \{ \mathrm{true} \} $ 
$\mathbfcal{H}_\mathrm{r} \rightarrow \{ \mathrm{true} \}$
and 
%$[\mathbfcal{H}_\mathrm{c} \sqcup_3 \tilde{\mathbfcal{H}}_\mathrm{r}] \rightarrow \{ \mathrm{fake} \} $
$\tilde{\mathbfcal{H}}_\mathrm{r} \rightarrow \{ \mathrm{fake} \}$.

\subsection{Adversarial Network Architecture}
\begin{figure}[tb!]
\centering
\includegraphics[width=\columnwidth]{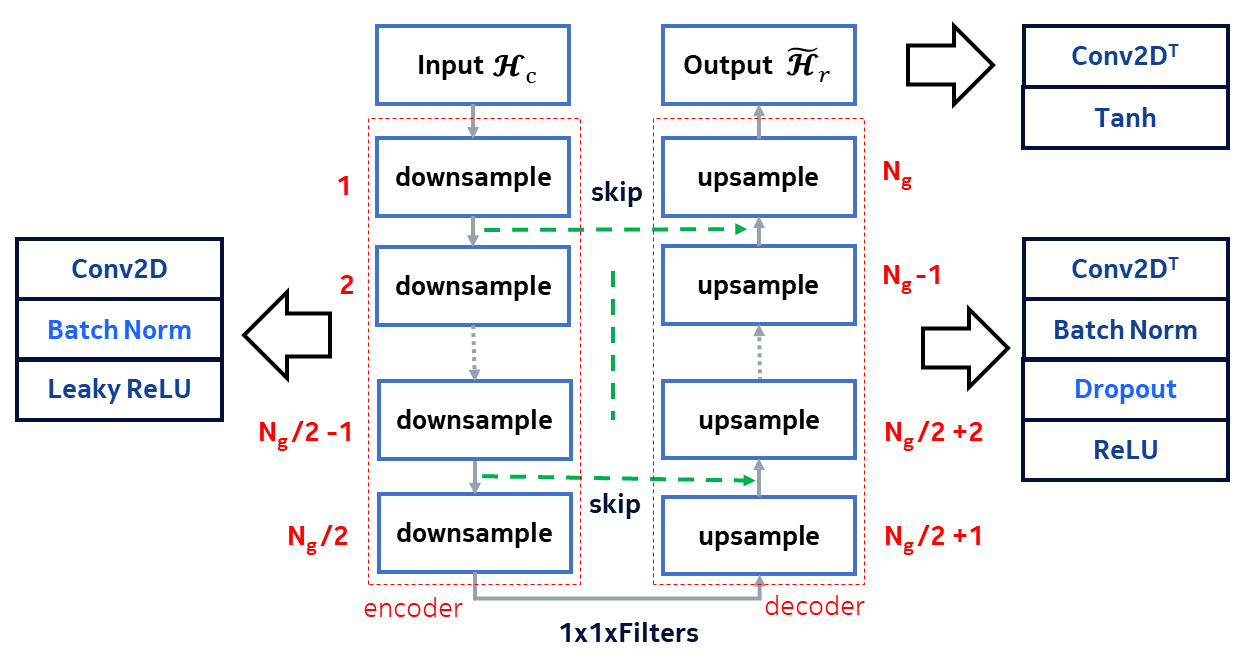}
\caption{U-Net architecture deployed as the generator including encoder and decoder pipeline and numbering for skip connections.}
\label{fig:u-net}
\end{figure}

Figure~\ref{fig:cgan} shows the interconnection between generator and discriminator NNs for the adversarial training. 
Here, the generator NN consists of an U-shaped deep NN (U-Net) which has two paths for the flow of information between blocks: the encoder-decoder path and the skip connections path, see Figure~\ref{fig:u-net}. 
The discriminator NN consists of a Patch-NN~\cite{17IsolaPix} where the input is reduced to a patch of arbitrary size; then, each coefficient of the patch is classified as real or fake. %, see Figure~\ref{fig:patch-net}. 

% In the encoder-decoder path, the input signal is sequentially encoded by $N_g/2$ downsampling blocks and decoded by the following $N_g/2$ upsampling blocks, where $N_g$ is the total number of processing blocks. 
%The skip connections path happens between blocks $n$ and $N_g-n$, where $n=[1:N_g]$, which provide more information to the decoder block~\cite{17IsolaPix}. 

%\begin{figure}[tb!]
%\centering
%\includegraphics[width=0.6\columnwidth]{figs/draws-discriminator.png}
%\caption{Patch-Net architecture deployed for the discriminator.}
%\label{fig:patch-net}
%\end{figure}

Figure~\ref{fig:u-net} shows the U-Net architecture employed for the generator, the $N_g/2$ downsample blocks for the encoder and the $N_g/2$ upsample blocks for the decoder, where $N_g$ is the total number of processing blocks. 
%According to design choices, each block can apply padding to its input $\mathbfcal{X}_g \in \mathbb{R}^{M_g \times P_g \times Q_g}$, where $g=[1:N_g]$ and $M_g, ~P_g,~ Q_g$ correspond to the input tensor dimensions. 
%The padding operation concatenates $\mathbf{0}$ vectors at columns and rows dimensions according to the size of the 
%filter $\mathbfcal{W}_g \in \mathbb{R}^{F_{g,1} \times F_{g,2} \times Q_g \times F_{g,3}}$ in the convolution operation,
%where $F_{g,1}, ~F_{g,2}$ corresponds to the number of rows and columns in the filter and $F_{g,3}$ is the number of filters in the $g^\mathrm{th}$ block. 
%The padded input is $\mathbfcal{X}_g' \in \mathbb{R}^{M_g+\floor{F_{g,1}/2} \times P_g\floor{F_{g,2}/2} \times Q_g}$.   
Each downsample block consists of one convolutional 2-dimensional layer (Conv2D), one batch normalization layer (BatchNorm), and a leaky rectifier linear unit (LeakyReLU) activation function, where $y = x$ for $x > 0$, and $y = 0.3x$ for $x<0$. 
%\begin{equation}
    %\mathbfcal{Y}_n = %\mathrm{LeakyReLU}(\mathrm{BatchNorm}(\mathrm{Padding}(\ma%thbfcal{Y}_{n-1}) \times \mathbfcal{W}_{g,n}))
%\end{equation}
Each upsample block consists of one transposed convolutional 2-dimensional layer (Conv2D$^T$), followed by BatchNorm and ReLU as activation function.
The skip connections path happens between the output of the $n^\mathrm{th}$ downsample block and the output of the $(N_g-n)^\mathrm{th}$ upsample block, where $n=[1, 2, \ldots, (N_g/2-1)]$. 
Those skip connections provide more information to the decoder block~\cite{17IsolaPix} since the input to each %$g^\mathrm{th}$ 
upsample block is the concatenation $\mathbfcal{X}_{(N_g-n+1)} = [\mathbfcal{Y}_n \sqcup_3 \mathbfcal{Y}_{(N_g-n)}]$, where $\mathbfcal{Y}_n$ is the output of the $n^\mathrm{th}$ block.
%, and $g=[(N_g/2+2), (N_g/2+3), \ldots, N_g]$. 

%In each convolutional layer, the number of filters $N_\mathrm{filter}$, their length and  how they are shifted (stride) are set such that, after the $N_b/2$ downsampling processing blocks, the input size is reduced to $[1 \times 1 \times N_\mathrm{filter}]$.  

%Figure~\ref{fig:patch-net} presents the discriminator NN, called Patch-NN. 
%The discriminator first reduces the dimensionality of the input signal, and classifies each output coefficient as real or fake. Hence, each output coefficient corresponds to a patch of the input signal.  
% For that, 
For the discriminator NN, we employ a Patch-NN~\cite{17IsolaPix}. First, downsampling blocks are used to reduce the dimensionality of the input signal to some patch of arbitrary size. 
Second, the patch is processed by a sequence of convolutional layers (Conv2D + BatchNorm + LeakyReLU and Conv2D + Linear).
Then, the discriminator is trained to classify each patch coefficient as real or fake. Implementations details are provided in Section~\ref{sec:results}.
%First, the discriminator concatenates the conditional input $\mathbfcal{H}_\mathrm{c}$ to the label $\mathbfcal{H}_r$ or to the generated channel $\tilde{\mathbfcal{H}}_\mathrm{r}$, forming, respectively, the real and fake classes. Then, the input is downsampled by the %it goes through 
%The architecture of the Patch-NN contains downsampling blocks which are followed by one zero padding layer, and one Conv2D with BatchNorm and LeakyReLU activation function. After one zero padding and one Conv2D layer, the discriminator provides its output which, during optimization, is further averaged and represented as a scalar value~\cite{17LedigPhoto}.

\subsection{Optimization with cGAN}
As shown in Figure~\ref{fig:cgan}, in cGAN there are two NNs playing a minmax game where the generator 
$G:\{ \mathbfcal{H}_\mathrm{c}\} \rightarrow \mathbfcal{H}_\mathrm{r}$ tries to fool the discriminator $D: \{ \tilde{\mathbfcal{H}}_\mathrm{r}\} \rightarrow \{\mathrm{true}\}$, and it is conditional because some prior knowledge is provided. Mathematically, the optimization objective of a cGAN has two terms  
\begin{equation}
G^* = \arg \min_{G} \max_{D}  \mathcal{L}_{\mathrm{cGAN}}(G,D) + \alpha \mathcal{L}_{\mathrm{L_2}},
    \label{eq:final-loss}
\end{equation}
where $\mathcal{L}_{\mathrm{cGAN}}(G,D)$ is the adversarial loss, $\mathcal{L}_{\mathrm{L_2}}$ is the $\mathrm{L_2}$ loss, and $\alpha$ is the weighting factor~\cite{17IsolaPix}. 
The adversarial loss is computed as
\begin{equation}
\begin{aligned}
\mathcal{L}_{\mathrm{cGAN}}(G,D) = & \mathbb{E}[\log D(\mathbfcal{H}_\mathrm{r})] + \\
& \mathbb{E}[\log (1-D(\tilde{\mathbfcal{H}}_\mathrm{r}))],
    \label{eq:cgan-loss}
\end{aligned}
\end{equation}
where the generator $G$ learns to map the input data $\mathbfcal{H}_\mathrm{c}$ to the output data $\mathbfcal{H}_\mathrm{r}$ such that 
$\tilde{\mathbfcal{H}}_\mathrm{r} = G^*(\mathbfcal{H}_\mathrm{c})$,
%$\rightarrow  \mathbfcal{H}_\mathrm{r}$
and the discriminator $D$ tries to recognize %keep up on recognizing 
the channels generated by $G$. In order to have the
generated
output wireless channels $\tilde{\mathbfcal{H}}_\mathrm{r}$ close to the wireless channel labels $\mathbfcal{H}_\mathrm{r}$, a weighted $L_2$ loss %$\mathcal{L}_{\mathrm{L1}}$ 
\begin{equation}
\mathcal{L}_{\mathrm{L_2}}(G) =  \mathbb{E}[\| \mathbfcal{H}_\mathrm{r}-G(\mathbfcal{H}_\mathrm{c}) \|_F]
    \label{eq:generator-l1}
\end{equation}
is included %on the generator loss function 
as a regularization term. In \cite{16Pathakcontext} a $L_2$ reconstruction loss is also proposed, but $0/1$ masks are used to restrict the $L_2$ loss only to the signal we aim to estimate. Note that, at the cGAN output, we have the conditional information as well as the desired signal. 
%constraining the regularization term only to the signal we aim to estimate - not considering the priors (or context). 
Hence, we study the feasibility of a reconstruction loss $\mathcal{L}_{\mathrm{rec}}$ defined as 
\begin{equation}
\mathcal{L}_{\mathrm{rec}} =  \mathbb{E}[\| \mathbf{H}_\mathrm{r}-\tilde{\mathbf{H}}_\mathrm{r} \|_F] +
\mathbb{E}[\| \mathbf{H}_\mathrm{LOC}-\tilde{\mathbf{H}}_\mathrm{LOC} \|_F],
    \label{eq:generator-l2}
\end{equation}
where the MSE of the target channel and the location matrix are considered. Then, $\mathcal{L}_{\mathrm{rec}}$ substitutes $\mathcal{L}_{\mathrm{L_2}}$ in Equation~\ref{eq:cgan-loss}.
%Therefore, the final optimization objective is 
%\begin{equation}
%G^* = \arg \min_{G} \max_{D}  \mathcal{L}_{\mathrm{cGAN}}(G,D) + \beta \mathcal{L}_{\mathrm{L_2}},
%    \label{eq:final-loss}
%\end{equation}
%where $\beta$ is the weighting factor. 

The generator and discriminator NNs are trained together in each epoch. For testing, or inference, only the generator architecture is used. Therefore, only knowledge of $\mathbfcal{H}_\mathrm{c}$ is needed. 
In practice, at inference time, we are able to estimate/predict a channel based on its location and the prior-knowledge provided by the UNNs. 

\section{Simulations and Results}
\label{sec:results}

Figure~\ref{fig:IlmProp} presents our propagation environment simulated at IlmProp where there are $3$ UEs used to derive  
%The three UEs in Figure~\ref{fig:IlmProp} are used to derive
$\mathbfcal{H}_\mathrm{mes}^C$ for prior-CSI 
estimation by the UNNs, and the red square indicates our study area where $\mathbfcal{H}_\mathrm{sim}^C$ is collected for training the cGAN to recreate the CSIs. 
Table~\ref{tab:SymIlm} presents the simulation parameters set at IlmProp. A total of seven simulation campaigns were performed, changing the number of UEs as well as their trajectories, and including or removing scatters. %The campaign shown in Figure~\ref{fig:IlmProp} is used to derive $\mathbfcal{H}_\mathrm{mes}^C$ for prior-CSI 
%estimation by the UNNs. 
%The other six simulation campaigns lies in the red square area where $\mathbfcal{H}_\mathrm{sim}^C$ is collected for training the cGAN to recreate the CSIs. 
The dataset for prior-CSI estimation with UNNs has 384 channel samples, and the dataset for cGAN has 4492 channel samples. 
We use the normalized squared error (NSE) 
$\mathrm{NSE} = \frac{\|\mathbf{B} -\mathbf{\tilde{B}}  \|_F^2}{\|\mathbf{B}\|_F^2}$
as our performance metric for CSI recreation. 

\begin{table}[tb!]
    \centering
    \caption{Simulation parameters for IlmProp.}
    \label{tab:SymIlm}
    \resizebox{0.4\linewidth}{!}{%
    \begin{tabular}{|c|c|c|}
    \hline
         Parameter & $\mathbfcal{H}_\mathrm{mes}^C$ & $\mathbfcal{H}_\mathrm{sim}^C$ \\ \hline
         Carrier frequency & \multicolumn{2}{c|}{2.6~GHz} \\ \hline
         Bandwidth & \multicolumn{2}{c|}{20~MHz} \\ \hline
         $N_\mathrm{sub}$ & \multicolumn{2}{c|}{64}  \\ \hline
         UE velocity & $1$ m/s & $0.1$ to $5$ m/s \\ \hline
         $N_\mathrm{sp}$ & 128 & 74 to 174 \\ \hline
         Total of UEs & 3 & many \\ \hline
         $N_\mathrm{ant}$ & \multicolumn{2}{c|}{36} \\ \hline
    \end{tabular}}
\end{table}

\begin{table}[bt!]
\centering
\caption{Description of the U-Net deployed as generator NN.}
\label{tab:gen-desc}
\resizebox{0.8\linewidth}{!}{%
\begin{tabular}{|c|c|c|c|c|c|c|c|c|}
\hline
 & Block & $N_\mathrm{filter}$ & Filter size & Stride & Padding & BatchNorm & Dropout & Activation\\ \hline
1 & downsample & 64 & [3,3] & [1,1] & Yes & No & No & LeakyReLU \\ \hline
2 & downsample & 64 & [2,4] & [1,2] & Yes & Yes & No & LeakyReLU \\ \hline
3 & downsample & 64 & [2,4] & [1,2] & Yes & Yes & No & LeakyReLU \\ \hline 
4 & downsample & 64 & [2,4] & [1,2] & Yes & Yes & No & LeakyReLU \\ \hline
5 & downsample & 64 & [2,4] & [1,2] & Yes & Yes & No & LeakyReLU \\ \hline
6 & downsample & 128 & [4,4] & [1,1] & No & Yes & No & LeakyReLU \\ \hline
7 & upsample & 64 & [4,4] & [2,2] & No & Yes & Yes & ReLU \\ \hline 
8 & upsample & 64 & [2,4] & [1,2] & Yes & Yes & Yes & ReLU \\ \hline
9 & upsample & 64 & [2,4] & [1,2] & Yes & Yes & Yes & ReLU\\ \hline
10 & upsample & 64 & [2,4] & [1,2] & Yes & Yes & No & ReLU\\ \hline
11 & upsample & 64 & [2,4] & [1,2] & Yes & Yes & No & ReLU \\ \hline
12 & upsample & 64 & [3,3] & [1,1] & Yes & Yes & No & ReLU \\ \hline
13 & output & 73 & [3,3] & [1,1] & Yes & No & No & TanH \\ \hline
\end{tabular}}
\end{table}

\begin{table}[bt!]
\centering
\caption{Description of the Patch-Net deployed as discriminator NN.}
\label{tab:dis-desc}
\resizebox{0.8\linewidth}{!}{%
\begin{tabular}{|c|c|c|c|c|c|c|}
\hline
 Block & $N_\mathrm{filter}$ & Filter size & Stride & Padding & BatchNorm & Activation \\ \hline
 downsample & 128 & [3,3] & [1,1] & Yes & Yes & LeakyReLU \\ \hline
 downsample & 128 & [2,4] & [1,2] & Yes & Yes & LeakyReLU \\ \hline
 zero padding 2D & - & - & - & Yes & - & -  \\ \hline
 Conv2D & 256 & [3,3] & [1,1] & No & Yes & LeakyReLU \\ \hline
 zero padding 2D & - & - & - & Yes & - & - \\ \hline
 Conv2D & 1 & [3,3] & [1,1] & No & No & Linear \\ \hline
 \end{tabular}}
\end{table}

\begin{figure}[tb!]
    \centering
    \includegraphics[width=\columnwidth]{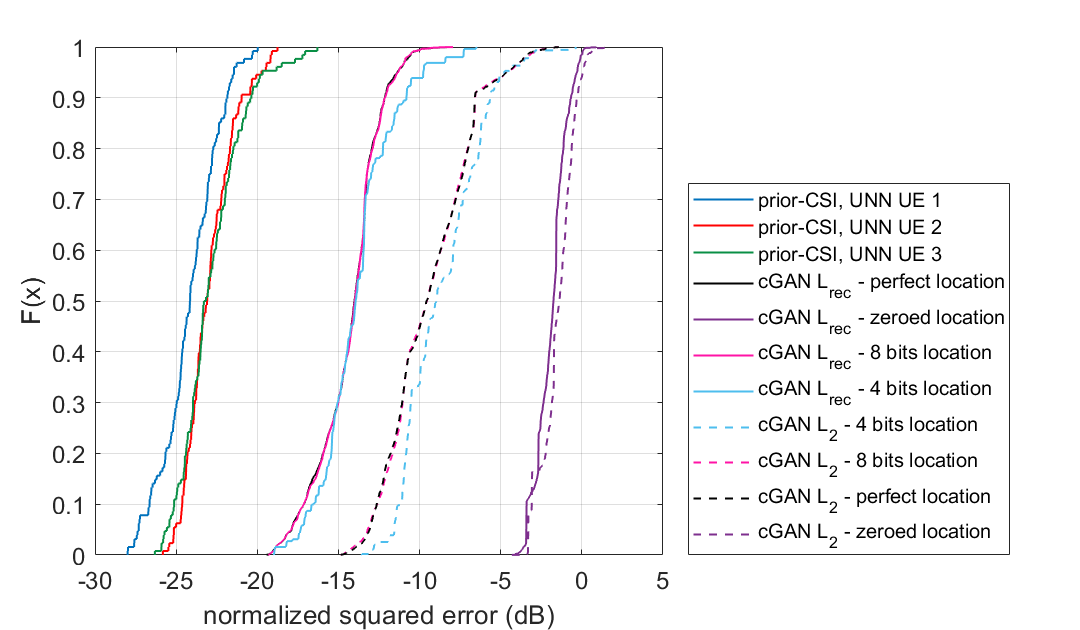}
    \caption{Cumulative distribution function of the normalized squared error for UNN and cGAN results for CSI recreation. The cGAN trained with $\mathcal{L}_{\mathrm{rec}}$ ($\mathrm{L}_\mathrm{rec}$) has best performance.}
    \label{fig:results}
\end{figure}

First, we define a UNN structure $P$ that is used to derive the optimal weights for all the three UEs in Figure~\ref{fig:IlmProp}. 
The $\mathbfcal{H}_\mathrm{mes}^C$ for each UE has $N_\mathrm{sp} = 128$ time snapshots and we design a UNN capable to estimate chunks of $64$ time snapshots. Therefore, we need to derive $2$ sets of trainable parameters $\mathbfcal{K}^*$ for each UE. In total, 
%Despite we use the same layer structure for all UEs, there is a specific set of trainable parameters $\mathbfcal{K}^*$ for each 64 time snapshots collected from a single UE trajectory. Therefore, a total of 
there are $6$ different UNN set of weights  $\mathbfcal{K}^*$ to estimate all the prior-CSIs. 
%, such that $\mathbfcal{H}_{\mathrm{p},1} = P(\mathbfcal{K}^*_1, \mathbfcal{Z}_0)$,  $\mathbfcal{H}_{\mathrm{p},2} = P(\mathbfcal{K}^*_2, \mathbfcal{Z}_0)$, and $\mathbfcal{H}_{\mathrm{p},3} = P(\mathbfcal{K}^*_3, \mathbfcal{Z}_0)$ are the channels estimated for UE $1$, $2$ and $3$, respectively. 
For the UNN mapping structure $P$, we choose to have four inner layers, each with $k_{1:L-2}=64$ filters and both upsampling matrices, $\mathbf{A}_l$ and $\mathbf{C}_l$, activated in all inner layers. 
In addition, we use one pre-output layer with $k_{L-1} = 64$ filters and one output layer with $k_L = 72$ convolutional filters. Therefore, there are $L=6$ layers and the random noise seed $\mathbfcal{Z}_0~\in~\mathbb{R}^{4 \times 4 \times 64}$ is drawn from a uniform distribution as $U(-0.15, +0.15)$ where $k_0=64$. 
After setting the UNN structure, the trainable parameters $\mathbfcal{K}$ are initialized from random values and $I=25000$ gradient updates are performed to
%we perform  $I=25000$ gradient updates and 
find the best $\mathbfcal{K}^*$ for each UE, separately. 
Our design choices for $k$ and $I$, the number of filters in each layer and the number of iterations respectively, 
were taken according to the estimated SNR at the UNN output. 
The UNN output SNR had to be at least the same as the SNR of the measured channel. 
Figure~\ref{fig:results} shows the cumulative distribution function (CDF) of the NSE for the UNN-estimator in blue, red and green curves for the $3$ UEs with $128$ time snapshots each, where $\mathrm{SNR}=20$~dB for the measured channel. 
%Here, $\mathrm{SNR}=20$~dB for the measured channel and the SNR of the prior-CSI estimation with UNN are $\mathrm{SNR}_{\mathbfcal{H}_{\mathrm{p},1}} = 23.05$~dB,
%$\mathrm{SNR}_{\mathbfcal{H}_{\mathrm{p},2}} = 22.43$~dB, and
%$\mathrm{SNR}_{\mathbfcal{H}_{\mathrm{p},3}} = 23.64$~dB. 
The presented UNN architecture is capable to recreate a sequence of $64$ time snapshots collected from a single UE. This structure contains $25,728$ trainable parameters which correspond to  $17.45\% $ of the coefficients in a channel measurement 
$\mathbfcal{H}_\mathrm{mes}^C~\in~\mathbb{C}^{64 \times 64 \times 36}$.
Therefore, UNNs provide means to compress the CSI with a high probability of about $90\%$ to recover it with better SNR than the measured one. 

After estimating the channels for the three UEs, we have a pool of $384$ prior-CSIs together with their location
$\mathbf{\Gamma}_\mathrm{p}$.
%$\mathrm{\textbf{LOC}}_\mathrm{p}$. 
Here, we use the location coordinates provided by the IlmProp simulator. 
In a practical implementation, we could derive the location from the estimated prior-CSIs by Unitary Tensor ESPRIT~\cite{08HaardtTensor}, for instance. 
A sub-set of $S=3$ CSI-location pairs are selected according to their minimum Euclidean distance to the location
$\mathbf{\Gamma}_\mathrm{r}$
%$\mathrm{\textbf{LOC}}_\mathrm{r}$ 
where we aim to recreate the CSI. 
Hence, the input to our cGAN is $\mathbfcal{H}_c \in \mathbb{R}^{4 \times 64 \times 73}$, where $64$ is the number of subcarriers. 
%From the simulations of $\mathbfcal{H}_\mathrm{sim}^C$, 
%we derive a total of $2812$ target CSI-location pairs $\{ \mathbfcal{H}_\mathrm{r}, \mathrm{\textbf{LOC}}_\mathrm{r}\}$ to be reconstructed by the cGAN.
The architecture details of our generator and discriminator are presented in Table~\ref{tab:gen-desc} 
and Table~\ref{tab:dis-desc}, respectively.  
The adversarial training runs for $150$ epochs with $60\%$ of the simulated channels used for training and $40\%$ for testing. 
Figure~\ref{fig:results} presents the CDF curves of the NSE  of the channels estimated by the cGAN with prior-knowledge and desired location as input. 
The black curves are the NSE for the cGANs trained with $\mathbf{H}_\mathrm{LOC}$ with infinity precision, where the full line and dashed line are for the cGANs trained with 
$\mathcal{L}_{\mathrm{rec}}$ and $\mathcal{L}_{\mathrm{L_2}}$ as regularization term, respectively.  
As we can see in Figure~\ref{fig:results}, the cGAN constrained to $\mathcal{L}_{\mathrm{rec}}$ achieves better performance with a $90\%$ NSE of about $-12.5$~dB. 

Regarding state of the art, our approach provides much higher performance if compared to the $6~$dB reported in ~\cite{21RatnanFadeNet}. Moreover, our cGAN architecture is less complex since we just use $13$ layers for the U-net at the generator while ~\cite{21RatnanFadeNet} has reported $28$ layers to process images of the environment map and output the wireless channel. The generator described in Table~\ref{tab:gen-desc} has a total of $789,786$ trainable parameters. Our cGAN takes about $6$~hours to train in a computer with $16$~GB of RAM and a GPU with $2$~GB of dedicated memory.  

At inference time, we test the sensitivity of the trained cGAN to a shut-down of $\mathbf{H}_\mathrm{LOC}$.
%the location matrix. 
We set $\mathbf{H}_\mathrm{LOC}=0$
%the location matrix to zero 
and compute the NSE at the cGAN output. This sensitivity curve is plotted in purple at Figure~\ref{fig:results}. 
The cGAN performance decays to nearly $0$~dB, which inform us that the architecture is not neglecting the location matrix.
Next, we test the cGAN sensitivity to location quantization errors. The locations are quantized using a uniform quantizer with $8$ and $4$ bits, their NSE are plotted in pink and cyan at Figure~\ref{fig:results}. 
There is very little performance loss even for $4$ bits location quantization, which introduces errors in the range of $1$ to $4$ times the channel wavelength. 
The cGAN is robust to those errors because it is closely modeling the channels statistical distribution. 
Our analyzes of the dataset has shown that the channels are mostly line of sight (LOS) and frequency-flat for the defined study area. 
Therefore, the good statistical behavior of the dataset makes the CSI-recreation task quite easy for the cGAN. Such robustness against location errors
%to errors in location 
may not be experienced in non-line of sight scenarios.

\section{Conclusion}
\label{sec:conclusion}

In this paper we propose to combine UNNs with cGAN to reconstruct wireless channels within an area in the propagation environment. The channel is reconstructed based on prior-CSIs from UNNs and the location where we aim to reconstruct the channel. Our method performs better than state of the art solutions, is much less complex and requires only some hours of training. Our results show that the cGAN performs well in a propagation scenario with mostly LOS channels. 
%For future work, we aim to apply this setup to non-line of sight scenarios where we may have more difficulty to train the cGAN as the channels might be very different in a centimeter range distance. 
Future work is related to further evaluation of quantization effects as well as advanced methods to improve the reconstructed CSI beyond this first ML inference.

\section*{Acknowledgement}
\small{This research was partly funded by German Ministry of Education and Research (BMBF) under grant 16KIS1184 (FunKI).}

\bibliographystyle{IEEEtran}
\bibliography{myREFfile}
\end{document}